# Mitigating Spreadsheet Risk in Complex Multi-Dimensional Models in Excel

Steve Litt
Indigo Sun, Inc.
36 Yardley Court Glen Rock, NJ 07452
sj@rumfox.com

**ABSTRACT**

*Microsoft Excel is the most ubiquitous analytical tool ever built. Companies around the world leverage it for its power, flexibility and ease of use. However, spreadsheets are manually intensive and prone to error, making it difficult for companies to control spreadsheet risk. The following solution is designed to mitigate spreadsheet risk for a set of problems commonly addressed in a spreadsheet defined as "complex multi-dimensional models". "Complex" referring to certain types of applications that require functionality such as sophisticated algorithms, challenging hierarchies and database write-back (i.e. planning, forecasting, etc.) and "multi-dimensional" referring to providing capabilities such as reporting, data input forms and ad hoc analysis on the different attributes associated with the resulting model. The solution is defined as a "PivotModel" because it works similarly to a PivotTable but is designed to leverage the robust capabilities of the Microsoft Excel platform.*

## 1. MITIGATING SPREADSHEET RISK IN COMPLEX MULTI-DIMENSONAL MODELS

Companies around the world leverage Microsoft Excel for its power, flexibility and ease of use. There are many different problems it addresses. One common problem is "complex multi-dimensional" models. The following is an example of a complex multi-dimensional model showcasing a P&L for a fictitious global Lighting Company.

| Quality Lighting Corp | | | | |
|---|---|---|---|---|
| Income Statement | | | | |
| Total Company | | | | |
| Total Products | | | | |
| Qtr1 | | | | |
| | Act/Fcst | Budget | $Var | %Var |
| Total sales | $ 228,973.89 | $ 230,259.35 | $ (1,285.46) | -0.6% |
| Discounts and allowances | $ 86.33 | $ 85.40 | $ 0.93 | 1.1% |
| Net sales | $ 228,887.56 | $ 230,173.95 | $ (1,286.39) | -0.6% |
| Standard cost of sales | $ 163,194.13 | $ 166,008.68 | $ (2,814.54) | -1.7% |
| Manufacturing Variances | $ 2,261.95 | $ 2,243.25 | $ 18.70 | 0.8% |
| Other Adjustments | $ 1,071.63 | $ 1,083.72 | $ (12.10) | -1.1% |
| Total cost of sales | $ 166,527.71 | $ 169,335.65 | $ (2,807.94) | -1.7% |
| Gross profit | $ 62,359.85 | $ 60,838.31 | $ 1,521.55 | 2.4% |
| Engineering | $ 10,346.88 | $ 10,312.56 | $ 34.32 | 0.3% |
| Research & development | $ 3,609.29 | $ 3,617.21 | $ (7.92) | -0.2% |
| General & administrative | $ 12,711.58 | $ 12,773.72 | $ (62.14) | -0.5% |
| Sales & marketing | $ 12,301.16 | $ 12,176.29 | $ 124.87 | 1.0% |
| Total operating expenses | $ 38,968.90 | $ 38,879.78 | $ 89.12 | 0.2% |
| Income from operations | $ 23,390.95 | $ 21,958.52 | $ 1,432.43 | 6.1% |

| Quality Lighting Corp | |
|---|---|
| **Offices** | **Products** |
| North | Commercial |
| West | Energy Savings |
| Central | LED Based |
| South | Outdoor |
|  Domestic |  Total Products |
| Europe | |
| Asia Pacific | |
|  International | |
|  Total Company | |
| **9** | **5** |

Figure 1: Sample P&L with Offices and Products

There are a few areas in this example that help define "complex". First is the reporting structure – notice the P&L rollup is not symmetrical, but instead ragged and unique to this company's reporting needs. Next, it is evident that the Budget column has a need for users to manually input data. Also, for line items such as "General & administrative" or "Sales & marketing", the detail costs might not be known for each office and product. For instance, this might be a cost only associated with Corporate and therefore requires an algorithm to allocate the cost down to each office and product to determine profitability.

The next component to this example helps define "multi-dimensional". Figure 1 is a report that the Quality Lighting Corporation wants to create for all offices and products. In this report, Accounts are in the rows, Scenarios are in the columns and the report is run for Quarter One. Therefore, counting totals and subtotals, there are nine (9) offices and five (5) products. Since PivotTables are not designed to support this complex problem, users resort to building out the combination, 45 (9x5) spreadsheets to support this model.

Also, these 45 spreadsheets are just for one (1) report. If the users want another type of report (i.e. P&L over Time) then that requires another 45 spreadsheets. Also, if this company adds another office and product, we are now up to 60 (10x6) spreadsheets to support this view.

If we introduce Time (i.e. Qtr1, Qtr2, Qtr3, Qtr4 and Year) to this model, we potentially need 12,600 cells (ACCTS (14) x SCENARIO (4) x ORG (9) x PRODUCT (5) x TIME (5)) to calculate all the combinations of this model. And, sans allocations etc., 1,728 cells (ACCTS (9) x SCENARIO (2) x ORG (6) x PRODUCT (4) x TIME (4)) or 14% of this model is data. The rest, 10,872 or 86%, of this model are calculations. Making sure these calculations are correct is incredibly difficult and this is a very simple example. …Imagine a model that contains 10,000,000 ($10^7$) cells?

Given this magnitude of potential calculations, it is easy to surmise that 9 out 10 spreadsheets have potential errors [Panko, 1992] and the consequences can be extreme. There are numerous stories where spreadsheet errors have caused companies huge financial losses.

## 2. CURRENT STATE OF THE ART OF SOLUTIONS

The current method of solving these types of problems is to leverage Excel's row, column, worksheet and workbook paradigm to build the desired multi-dimensional model. The model design is predicated on being able to address the permutations of the members associated with each dimension. In the example above, we are dealing with 12,600 cells.

When creating a model, users are faced with the need to manually develop and enhance their spreadsheets. These labor-intensive solutions are prone to error and introduce the potential for spreadsheet risk. There are techniques that leverage more sophisticated methods [Bartholomew, 2016] like range names, array formulas, lookup functions and other techniques to help mitigate spreadsheet risk. Implementing these solutions involves a more experienced spreadsheet user but still have the potential for error.

Once a spreadsheet is complete and if/when an error occurs, it is extremely difficult to diagnose. As well, spreadsheets fall short in providing documentation that meet regulatory requirements. There are several tools in the market today [Schalkwijk, 2015] designed to facilitate a user's effort it recognizing incorrect calculation logic to mitigate risk. These solutions, however, are utilized after the model is built. They reduce risk by searching through models and isolating on areas that meet criteria of a list of common problems they are aware of. Our solution is designed to programmatically help in the initial model development, thereby mitigating, if not eliminating, the "human error" risks.

## 3. NEW ALTERNATIVE SOLUTION

The following is a new method in addressing complex multi-dimensional modeling in Excel. It is a programmatic approach to calculation development that mitigates the potential of human error and consequently, spreadsheet risk.

We call our solution a "PivotModel". The reason we use the term PivotModel is to differentiate it as a more robust version of a PivotTable. However, the PivotModel is an all-in-one solution with characteristics that are also similar to Microsoft's Data Models and Power Pivot.

For example, a Data Model provides the capability to integrate data from multiple tables, effectively building a relational data source inside an Excel workbook. The PivotModel acts as a meta data container that allows users to load disparate data from multiple tables without the need for SQL. The result is an excel-based workbook indexed to provide multi-dimensional slice and dice capability like a PivotTable.

It is similar to Power Pivot in that it provides a language to create calculations. For a PivotModel, the language is the functions and calculation capabilities found in Excel. Users find it easy to get up to speed on the rule logic because they are familiar with native Excel. For Power Pivot, it is a specific language called DAX that is relational-like and used to produce calculations.

### 3.1 Technical Overview

The PivotModel can best be described as a MOLAP solution in that calculations are pre-generated for fast query performance. It also handles complex calculations and read/write access to support Reporting, Analysis, Modeling and Planning applications.

ROLAP solutions like Tableau and others differ significantly in that the common data storage is relational and they do not pre-compute and store information. The ROLAP solutions are designed to handle Reporting and Analysis.

The PivotModel is designed to complement Microsoft's existing toolset. For instance, the PivotTable handles rack'em stack'em aggregation to support analysis and data visualization. There are, however, a plethora of more complex business problems that these types of solutions don't address.

For instance, consider an application that requires users to compare price-volume-mix to better understand product variances. Or, an application that handles reciprocal allocations that perform simultaneous equations to distribute indirect costs to profit centers. Or, a solution that models a leverage buyout (LBO) comparing internal rates of return from different what-if scenarios. These are the areas where Excel really shines.

By creating a solution that resides on top of the Excel technology, we not only tap into the power of Excel, but amplify its results by simplifying rule logic, ensuring calculation accuracy and providing multi-dimensional slicing and dicing capabilities.

There are three aspects to developing a PivotModel: Building a Model Structure; Creating Rule Logic; and Loading Data. Once complete, the user develops Views and/or Input Forms from the PivotModel for reporting, analysis, modeling and planning. This method differs from other technologies in that it approaches analysis from a "Top-Down" perspective instead of "Bottoms Up".
Technologies like PivotTables, Power Pivot, Tableau, etc. focus on gathering data and merging it together resulting in a table that houses all the data and meta data in which to perform analysis. Thus, the concept of Bottoms Up because analysis is driven by the resultant table produced. Whereas with a PivotModel, users create the desired analysis first and then procure the required data to support it. This Top Down approach permits flexibility in modifying meta data and calculations as well as provides methods to integrate data not easily performed with SQL.

### 3.2 Building a Model Structure

The Model Structure includes dimensions, members of the dimensions and the member hierarchies associated with each dimension. For instance, utilizing our Lighting Company example, we have five dimensions: Accounts, Time, Product, Organization and Scenario. The following represents the members in that model structure.

|   | A | B | C | D | E | F | G | H | I | J | K | L |
|---|---|---|---|---|---|---|---|---|---|---|---|---|
| 1 |   |   |   |   |   |   |   |   |   |   |   |   |
| 2 |   | ACCTS |   |   |   | TIME |   | PRODUCT |   | ORG |   | SCENARIO |
| 3 |   | Total sales |   |   |   | Qtr1 |   | Commercial |   | North |   | Actuals |
| 4 |   | Discounts and allowances |   |   |   | Qtr2 |   | Energy Savings |   | South |   | Budget |
| 5 |   | Net sales |   |   |   | Qtr3 |   | LED Based |   | West |   | $Var |
| 6 |   | Standard cost of sales |   |   |   | Qtr4 |   | Outdoor |   | Central |   | %Var |
| 7 |   | Manufacturing Variances |   |   |   | Year |   | Total PRODUCT |   | Total Domestic |   |   |
| 8 |   | Other Adjustments |   |   |   |   |   |   |   | Europe |   |   |
| 9 |   | Total cost of sales |   |   |   |   |   |   |   | Asia Pacific |   |   |
| 10 |   | Gross profit |   |   |   |   |   |   |   | Total Europe/ASiaPac |   |   |
| 11 |   | Engineering |   |   |   |   |   |   |   | Total ORG |   |   |
| 12 |   | Research & development |   |   |   |   |   |   |   |   |   |   |
| 13 |   | General & administrative |   |   |   |   |   |   |   |   |   |   |
| 14 |   | Sales & marketing |   |   |   |   |   |   |   |   |   |   |
| 15 |   | Total operating expenses |   |   |   |   |   |   |   |   |   |   |
| 16 |   | Income from operations |   |   |   |   |   |   |   |   |   |   |
| 17 |   |   |   |   |   |   |   |   |   |   |   | Total |
| 18 |   | 14 |   |   |   | 5 |   | 5 |   | 9 |   | 4 | 12600 |

Figure 2: Sample Dimensions and Members

Therefore, there are five dimensions in this model, Accounts has ten (14) members, Time has five (5) members, Product has five (5) members, Org has nine (9) members and Scenario has four (4) members. The number of cells in this simple model with data and calculations is 12,600 (14x5x5x9x4). Each of these dimensions has a hierarchy. For example, one could drill down on each of these dimensions to detail. For example on TIME, we could drill into "Year" which is comprised of Qtr1 thru Qtr4.

|   |   |   |
|---|---|---|
| 8 |   |   |
| 9 | TIME | Collapsed |
| 14 | Year |   |
| 15 |   |   |
| 16 | TIME | Expanded |
| 17 | Qtr1 |   |
| 18 | Qtr2 |   |
| 19 | Qtr3 |   |
| 20 | Qtr4 |   |
| 21 | Year |   |

Figure 3: Time Dimension Collapsed and Expanded

### 3.3 Creating Rule Logic

The following depicts the rules associated with the Accounts dimension of this model. The dimension and its members are in a hierarchy expressed in Column A. In Column B resides the data with Excel Rule

Logic. And, in Column C, the Excel cell logic is displayed. Therefore, cell B5 (9,900.58) is computed as B3–B4.

|   | A | B | C |
|---|---|---|---|
| 1 |   |   |   |
| 2 |   | DATA | Excel Calc |
| 3 | Total sales | 9,904.06 |   |
| 4 | Discounts and allowances | 3.48 |   |
| 5 | Net sales | 9,900.58 | =B3-B4 |
| 6 | Standard cost of sales | 6,616.63 |   |
| 7 | Manufacturing Variances | 93.74 |   |
| 8 | Other Adjustments | 44.23 |   |
| 9 | Total cost of sales | 6,754.60 | =SUM(B6:B8) |
| 10 | Gross profit | 3,145.98 | =B5-B9 |
| 11 | Engineering | 447.21 |   |
| 12 | Research & development | 143.23 |   |
| 13 | General & administrative | 551.71 |   |
| 14 | Sales & marketing | 514.10 |   |
| 15 | Total operating expenses | 1,656.26 | =SUM(B11:B14) |
| 16 | Income from operations | 1,489.71 | =B10-B15 |
| 17 | Income % of Total Sales | 15% | =B16/B3 |

Figure 4: Accounts Dimension with Data and Rules

In the Excel paradigm, every calculation references a cell or range of cells. Therefore, when reviewing the calculation it will refer to the Column and Row coordinates (e.g. B5 or R5C2). When dealing with large amounts of cells with logic, there is the potential of users mistyping or referencing the wrong cells in Excel.

From a modeling perspective, it is better to refer to the calculation in business terms. In this case, the business logic associated with this calculation is really "Net Sales", which equals "Total Sales" minus "Discounts and Allowances". Our application creates Rules instead of cell logic making the model easier to understand and simpler for users to ensure the model is computing accurately. Here are the Rules (12) associated with the Lighting Company's PivotModel.

| Name | Rules | Type |
|---|---|---|
| Main - Calculations |   | PF |
| 1 - ORG - Domestic | ={North}+{West}+{Central}+{South} | BI |
| 2 - ORG - International | ={Europe}+{Asia Pacific} | BI |
| 3 - ORG - Total Company | ={Domestic}+{International} | BI |
| 4 - PRODUCT - Total Products | ={Commercial}+{Energy Savings}+{LED Based}+{Outdoor} | BI |
| 5 - TIME - Year | ={Qtr1}+{Qtr2}+{Qtr3}+{Qtr4} | BI |
| 6 - ACCTS - Net sales | ={Total sales}-{Discounts and allowances} | BI |
| 7 - ACCTS - Total cost of sales | =SUM({Standard cost of sales},{Manufacturing Variances... | BI |
| 8 - ACCTS - Gross profit | ={Net sales}-{Total cost of sales} | BI |
| 9 - ACCTS - Total operating expenses | =SUM({Engineering},{Research & development},{General... | BI |
| 10 - ACCTS - Income from operations | ={Gross profit}-{Total operating expenses} | BI |
| 11 - SCENARIO - $Var | =IFERROR({Act/Fcst}-{Budget},0) | BI |
| 12 - SCENARIO - %Var | =IFERROR({$Var}/{Act/Fcst},0) | BI |

Figure 5: Rule Logic in Application

Rules 6-10 in Figure 5 reflect the same logic as the Excel calculations example displayed in Figure 4. The Rule Editor acts as a calculation interpreter walking through the Rules in sequence and applying them to the PivotModel. The application utilizes Excel's extensive library of functions (i.e. VLOOKUP, etc.) with the ability to parameter drive the rules with dimension members and/or provide additional logic for calculations such as time intelligence. The Column Type refers to whether the line is a Standard or Custom Rule or a Folder or Sub-folder.

The major difference is the Rules represent logic as it pertains to every combination within the model. Therefore, "Net Sales" is computed for every PRODUCT, TIME, ORG and SCENARIO. This dramatically reduces spreadsheet risk because one rule refers to numerous calculations in the resulting Excel model. For instance, this particular Rule represents 900 calculated cells (PRODUCT (5) x TIME (5) x ORG (9) x SCENARIO (4))

It is important to note that this represents the maximum number of calculations for this formula. However, in an analytical model like this, in the other dimensions, for instance, "Year" in the TIME dimension, there is the potential of two calculations producing the same results. To clarify, in some situations, we could add up "Net Sales for the Year" (i.e. "Net Sales for Qtr1"+"Net Sales for Qtr2"+etc.) and it would equal the same results as if we computed "Net Sales" as the difference between "Total Sales for the Year" minus "Discounts and Allowances for the Year".

The scenario above produces the same results, however, in different situations, this might not be the case. Consequently, it is imperative that the right Rule take precedence or your model may produce incorrect results. Using our "Net Sales" and "Year" example, if these Rules took precedence in the calculation for the "Var %" member in the SCENARIO dimension, it would produce the wrong results because the Year rule would add up the %Var for all Qtrs which is incorrect.

The solution allows users to control the calculation precedence by providing the ability to control the sequence in which the calculation logic takes place. Using the standard Excel framework, users must determine the precedence of every calculation (12,600 cells) in their model. Anyone familiar with Excel knows how easy it is to err by dragging forward a rule (i.e. copy and paste) to additional cells when it is not the right calculation.

Excel is a very powerful and flexible solution. However, it is error prone and manually intensive because users have to perform repetitive rule logic and determine proper calculation precedence. Our application is designed to programmatically take care of this repetitive logic development and calculation precedence to mitigate spreadsheet risk.

The goal is to provide functionality that matches or exceeds the capabilities inherent in Excel. Therefore, at a minimum, anything you can do in Excel, you should be able to do in the solution. For instance,

- Filtering – being able to apply logic to specific members in the PivotModel (i.e. Only apply a calculation to specific members)
- Cross-Dimensions – being able to apply logic across dimensions (i.e. Each Product's Sales as a % to Total Sales for All Products)
- Time Intelligence – being able to create calculations for time (i.e. Year-to-Date, This Year vs Prior Year, etc)
- Calculate and Store – able to produce logic that results in data giving the user the ability to override the results (create a forecast from historical data and allow the user to override)

The final results reside completely in Excel. There is no multi-dimensional data source outside the Excel framework that gets queried and then displays the data in the spreadsheet.

### 3.4 Loading Data

Every model stores base level data. For instance, Column C in Figure 4 displays Excel rule logic for the cells in Column B. The other cells in Column B are considered Data.

The model structure acts as a container. Users can load data from multiple data sources to different parts of the PivotModel. For instance, we may have Actual and Budget data. The example below is Actual data from aback-end ERP in the following format, while the Budget data may come from other data sources.

| | A | B | C | D | E | F |
|---|---|---|---|---|---|---|
| 1 | ACCTS | SCENARIO | TIME | ORG | PRODUCT | Value |
| 2 | Cost of Sales | Actuals | Qtr1 | Asia Pacific | Outdoor | 6,602.56 |
| 3 | Cost of Sales | Actuals | Qtr1 | Europe | Outdoor | 6,626.30 |
| 4 | Cost of Sales | Actuals | Qtr1 | West | Outdoor | 6,585.05 |
| 5 | Cost of Sales | Actuals | Qtr1 | South | Outdoor | 6,877.30 |
| 6 | Cost of Sales | Actuals | Qtr1 | Central | Outdoor | 6,665.89 |
| 7 | Cost of Sales | Actuals | Qtr1 | North | Outdoor | 6,575.59 |
| 8 | Cost of Sales | Actuals | Qtr1 | Asia Pacific | LED Based | 7,020.86 |
| 9 | Cost of Sales | Actuals | Qtr1 | Europe | LED Based | 6,559.15 |
| 10 | Cost of Sales | Actuals | Qtr1 | West | LED Based | 6,934.92 |
| 11 | Cost of Sales | Actuals | Qtr1 | South | LED Based | 6,723.16 |
| 12 | Cost of Sales | Actuals | Qtr1 | Central | LED Based | 6,779.65 |
| 13 | Cost of Sales | Actuals | Qtr1 | North | LED Based | 6,956.21 |
| 14 | Cost of Sales | Actuals | Qtr1 | Asia Pacific | Energy Savings | 6,741.00 |
| 15 | Cost of Sales | Actuals | Qtr1 | Europe | Energy Savings | 6,663.97 |
| 16 | Cost of Sales | Actuals | Qtr1 | West | Energy Savings | 6,990.76 |
| 17 | Cost of Sales | Actuals | Qtr1 | South | Energy Savings | 7,148.46 |
| 18 | Cost of Sales | Actuals | Qtr1 | Central | Energy Savings | 6,838.62 |
| 19 | Cost of Sales | Actuals | Qtr1 | North | Energy Savings | 7,188.81 |
| 20 | Cost of Sales | Actuals | Qtr1 | Asia Pacific | Commercial | 6,597.92 |

Figure 6: Sample Load File for Actual Data from the Company ERP

Here is an example of another data load file type. This one is specific for Europe's Budget Data with the Accounts going across the columns. The data can be imported and exported in different formats.

| | A | B | C | D | E | F | G | H | I |
|---|---|---|---|---|---|---|---|---|---|
| 1 | SCENARIO | TIME | ORG | PRODUCT | Sales | Discounts and allowances | Cost of Sales | Manufacturing Variances | Other Adjustments |
| 2 | Budget | Qtr1 | Europe | Commercial | 9108.738986 | 3.449074244 | 7091.923907 | 91.48248473 | 45.57454258 |
| 3 | Budget | Qtr1 | Europe | Energy Savings | 9411.467163 | 3.758920712 | 6641.750863 | 97.35509751 | 45.11864966 |
| 4 | Budget | Qtr1 | Europe | LED Based | 9635.398421 | 3.692983947 | 7037.316718 | 96.01557864 | 45.48637086 |
| 5 | Budget | Qtr1 | Europe | Outdoor | 9866.786353 | 3.528756623 | 7183.396734 | 89.72451894 | 44.56970447 |
| 6 | Budget | Qtr2 | Europe | Commercial | 9841.565392 | 3.593072155 | 7118.463194 | 97.13626758 | 46.30703589 |
| 7 | Budget | Qtr2 | Europe | Energy Savings | 10037.09979 | 3.496472204 | 6606.107241 | 89.54798002 | 46.18606719 |
| 8 | Budget | Qtr2 | Europe | LED Based | 9613.146397 | 3.642690704 | 6975.576582 | 98.14827491 | 42.94386551 |
| 9 | Budget | Qtr2 | Europe | Outdoor | 9566.731302 | 3.671749628 | 6921.643558 | 91.00037981 | 44.31068596 |
| 10 | Budget | Qtr3 | Europe | Commercial | 9362.610195 | 3.53560235 | 7201.291727 | 95.48832235 | 46.30819488 |
| 11 | Budget | Qtr3 | Europe | Energy Savings | 9531.207687 | 3.576583843 | 6947.217344 | 90.5045327 | 43.26204336 |
| 12 | Budget | Qtr3 | Europe | LED Based | 9341.954244 | 3.742703733 | 6833.8511 | 90.73843317 | 47.03076267 |
| 13 | Budget | Qtr3 | Europe | Outdoor | 9690.730933 | 3.696396103 | 6700.8416 | 89.62534115 | 43.70483744 |
| 14 | Budget | Qtr4 | Europe | Commercial | 9767.0606 | 3.776939986 | 6679.99507 | 94.85694808 | 45.61505118 |
| 15 | Budget | Qtr4 | Europe | Energy Savings | 9148.44352 | 3.770450585 | 6808.857449 | 92.43427407 | 43.60567859 |
| 16 | Budget | Qtr4 | Europe | LED Based | 9554.792254 | 3.707764742 | 7081.413343 | 90.84821719 | 46.79817787 |
| 17 | Budget | Qtr4 | Europe | Outdoor | 10032.64548 | 3.454080212 | 7163.510058 | 89.31782528 | 46.57734516 |

Figure 7: Sample Data Load File for Europe's Budget Data

Both Figure 6 and Figure 7 are examples of different data file formats that a user can use to load data into the PivotModel. The PivotModel acts as a container where users load data from different disparate data sources. Data can be imported if it is associated with unique members of each dimension. Therefore, users can load data from multiple files. There is no need to consolidate all the data into one data set before viewing the data as required in a PivotTable. This reduces the need for advanced SQL skills to analyze information.

### 3.5 Views & Input Forms

The application provides reporting and ad hoc capabilities to view the multi-dimensional data once the PivotModel is complete. If the PivotModel resides on a universal drive, multiple users can access it for reporting and ad hoc analysis. The following is a sample view.

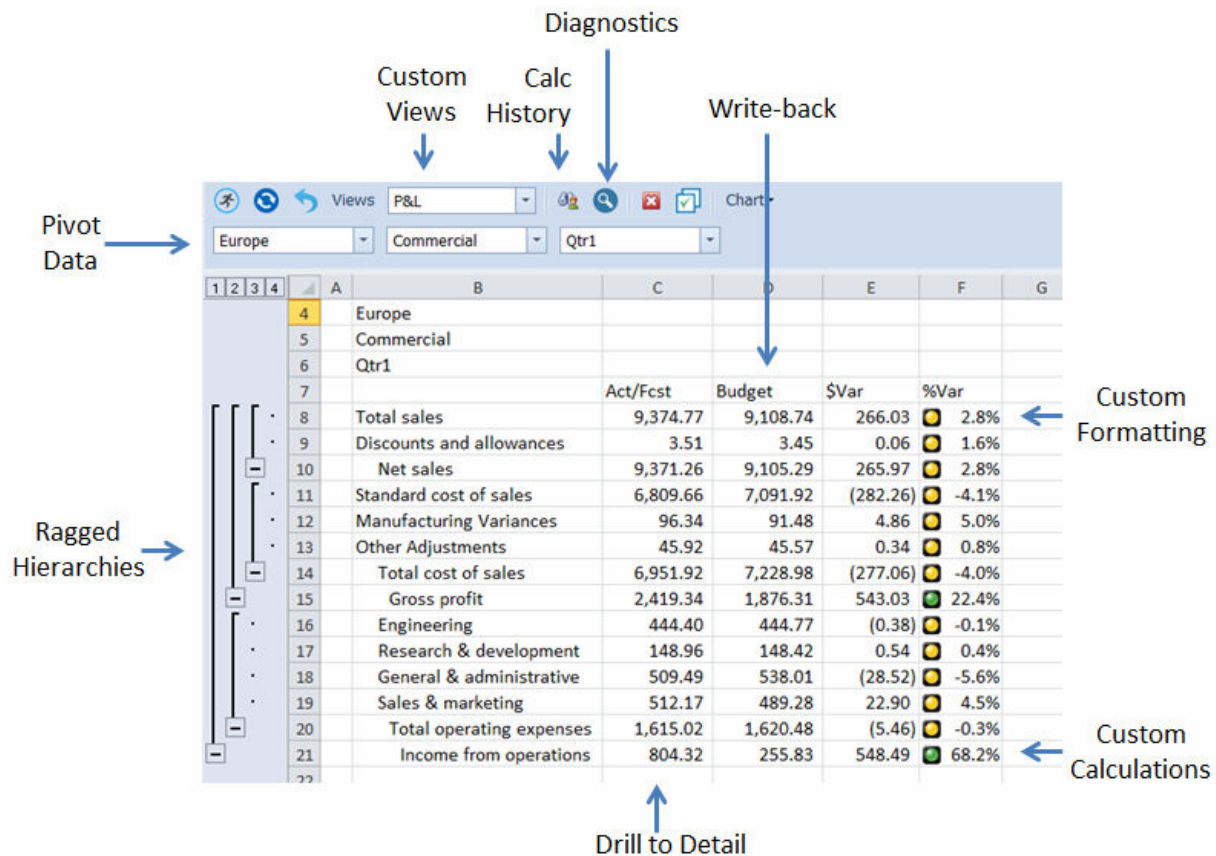

Figure 8: Results in Sample PivotModel View

**Figure 8 Captions**

- Custom Views - The solution provides developers with the ability to build custom views against the PivotModels. It supplies designers with menu controls to give users seamless access to different PivotModel reports.

- Write-Back (for Planning, Forecasting and What-if Analysis) - At the solution core is Excel. Therefore, users can write back any type of data (numeric, text, dates, url, etc.) to the PivotModel.

It is possible to design robust driver-based solutions for modeling. The following depicts how the PivotModel is accessed via their Excel front-end.

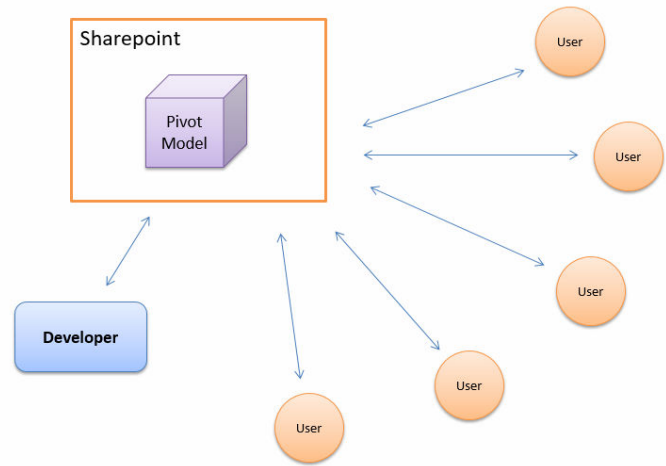

Once the PivotModel is developed, users can pull data to view results. But they also can "write-back" to the PivotModel. For instance, User1 is responsible for Business Unit 1. Corporate needs User1 to provide a Budget for next year. User1 brings up a view and then enters their budget and hits "send" and the data is written back to the PivotModel.

- Ragged Hierarchies (for Financial Reporting) - Users need the flexibility to build any type of financial or management report. The solution allows users to create reports the way users want to see them without constraints commonly found in tools like PivotTables and Tableau.

- Pivot Data (for Slicing and Dicing) -The solution allows users to perform multi-dimensional analysis to slice and dice and drill down into pivot table detail.

- Utilizes Excel capabilities - If you can do it in Excel, you can do it within the application. Leverage all of Excel's functionality - fonts, data types, themes, conditional formatting, functions, macros, add-ins, sql database access, PowerPivot. All these capabilities are available within the framework.

- Drill-down to Relational detail - If a PivotModel is developed off a relational table or PivotTable, the solution allows users to drill down with a focused query to the supporting table detail. These tables can be linked to outside data sources via traditional Excel functionality or third party tools that access proprietary data sources.

- Diagnostics - The solution allows users to trace through the PivotModel by the formulas that produced the result. The diagnostic tool gives users detail explanation on how each cell was computed. This is explained further in the next section.

- Link PivotModels to PivotModels - The application provides the ability to link data between different PivotModels. This simplifies development of complex business solutions. For instance, create consolidated budget pivot tables derived from a multitude of different departmental budgets.

### 3.6 Diagnostics & Documentation

The solution captures the interrelationships of the algorithms to provide methods in which to diagnose results. The following is a sample View for our Lighting Company. We are at a consolidated level and recognize a -0.6% for "% Var of Net Sales" and want to investigate the details.

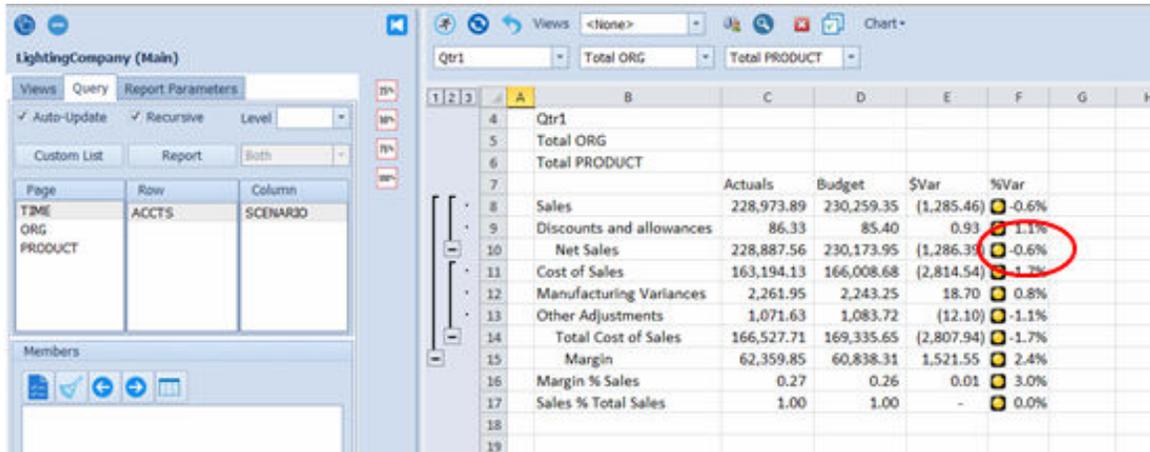

Figure 9: Sample PivotModel Query at Qtr1 for Total Org and Total Product

Double-clicking the cell drills into the detail taking the user to a separate worksheet to analyze the calculation logic that produced the results. The user can double-click on cells to step through the logic for each calculation to determine how it was derived. This capability is best described as a multi-dimensional Trace Precedence where users step through the business logic to determine how certain cells are computed.

The following shows a user drilling into the formula for %Var of Net Sales. First drilling into the numerator of the Net Sales' formula (Actual – Budget). Then drilling into Budget for Total Product, then drilling into Outdoor products to get to Total Organization, then drilling into Europe from the Total Europe/AsiaPac, then drilling into Europe's Net Sales to see the detail data supporting Net Sales: Sales – Discount Allowances.

|   | A | B | C | D | E | F | G | H | I |
|---|---|---|---|---|---|---|---|---|---|
| 1 | None | Level | SCENARIO | TIME | ORG | PRODUCT | ACCTS | Data | Rule |
| 2 |   | L1 | %Var | Qtr1 | Total ORG | Total PRODUCT | Net Sales | -0.00562 | IFERROR({$Var}/{Actuals},0) |
| 3 |   | L1.1 | $Var | Qtr1 | Total ORG | Total PRODUCT | Net Sales | -1,286.39090 | IFERROR({Actuals}-{Budget},0) |
| 4 |   | L1.2 | Actuals | Qtr1 | Total ORG | Total PRODUCT | Net Sales | 228,887.56337 | {Commercial}+{Energy Savings}+{LED Based}+{Outdoor} |
| 5 |   |   |   |   |   |   |   |   |   |
| 6 |   | L1.1 | $Var | Qtr1 | Total ORG | Total PRODUCT | Net Sales | -1,286.39090 | IFERROR({Actuals}-{Budget},0) |
| 7 |   | L1.1.1 | Actuals | Qtr1 | Total ORG | Total PRODUCT | Net Sales | 228,887.56337 | {Commercial}+{Energy Savings}+{LED Based}+{Outdoor} |
| 8 |   | L1.1.2 | Budget | Qtr1 | Total ORG | Total PRODUCT | Net Sales | 230,173.95426 | {Commercial}+{Energy Savings}+{LED Based}+{Outdoor} |
| 9 |   |   |   |   |   |   |   |   |   |
| 10 |   | L1.1.2 | Budget | Qtr1 | Total ORG | Total PRODUCT | Net Sales | 230,173.95426 | {Commercial}+{Energy Savings}+{LED Based}+{Outdoor} |
| 11 |   | L1.1.2.1 | Budget | Qtr1 | Total ORG | Commercial | Net Sales | 55,996.36510 | {Total Domestic}+{Total Europe/ASiaPac} |
| 12 |   | L1.1.2.2 | Budget | Qtr1 | Total ORG | Energy Savings | Net Sales | 58,053.42649 | {Total Domestic}+{Total Europe/ASiaPac} |
| 13 |   | L1.1.2.3 | Budget | Qtr1 | Total ORG | LED Based | Net Sales | 57,169.72327 | {Total Domestic}+{Total Europe/ASiaPac} |
| 14 |   | L1.1.2.4 | Budget | Qtr1 | Total ORG | Outdoor | Net Sales | 58,954.43941 | {Total Domestic}+{Total Europe/ASiaPac} |
| 15 |   |   |   |   |   |   |   |   |   |
| 16 |   | L1.1.2.4 | Budget | Qtr1 | Total ORG | Outdoor | Net Sales | 58,954.43941 | {Total Domestic}+{Total Europe/ASiaPac} |
| 17 |   | L1.1.2.4.1 | Budget | Qtr1 | Total Domestic | Outdoor | Net Sales | 39,135.88195 | {North}+{South}+{West}+{Central} |
| 18 |   | L1.1.2.4.2 | Budget | Qtr1 | Total Europe/ASiaPac | Outdoor | Net Sales | 19,818.55745 | {Asia Pacific}+{Europe} |
| 19 |   |   |   |   |   |   |   |   |   |
| 20 |   | L1.1.2.4.2 | Budget | Qtr1 | Total Europe/ASiaPac | Outdoor | Net Sales | 19,818.55745 | {Asia Pacific}+{Europe} |
| 21 |   | L1.1.2.4.2.1 | Budget | Qtr1 | Asia Pacific | Outdoor | Net Sales | 9,955.29986 | {Sales}-{Discounts and allowances} |
| 22 |   | L1.1.2.4.2.2 | Budget | Qtr1 | Europe | Outdoor | Net Sales | 9,863.25760 | {Sales}-{Discounts and allowances} |
| 23 |   |   |   |   |   |   |   |   |   |
| 24 |   | L1.1.2.4.2.2 | Budget | Qtr1 | Europe | Outdoor | Net Sales | 9,863.25760 | {Sales}-{Discounts and allowances} |
| 25 |   | L1.1.2.4.2.2.1 | Budget | Qtr1 | Europe | Outdoor | Sales | 9,866.78635 |   |
| 26 |   | L1.1.2.4.2.2.2 | Budget | Qtr1 | Europe | Outdoor | Discounts and allowances | 3.52876 |   |

Figure 10: Sample Drill Down on Net Sales for %Var

The solution also houses details of the model design to support internal Business and IT needs as well as regulatory requirements. The following are excerpts from the documentation.

Figure 11: Parent/Child Dimension Table and Rule Logic documentation

In summary, we mentioned 12,600 cells in which 1,728 cells (14%) were data and the remaining 10,872 cells (86%) were calculations. Since the application programmatically develops all the calculations, there are only 12 rules to manage the logic of 10,872 cells dramatically simplifying the development and management of complex multi-dimensional models.

## 4. THE SOLUTION COMPARED TO OTHERS

There are many solutions in the market, both from a design and product perspective, that are focused on issues associated with complex multi-dimensional modeling. For example, in "How do you know your spreadsheet is right?" [Bewig, 2005], there are step by step processes one can go through to minimize spreadsheet risk. Interestingly, in one of the diagrams, there is a "dependence graph" reflecting the concept of Rules. This approach, as well as others, addresses the problem in a similar manner by manufacturing a solution assuming the Excel Row, Column, Worksheet, Workbook paradigm as its baseline. The solution starts with a conceptual framework first (Build Model Structure, Create Rules, and Load Data) then develops the model leveraging the Excel paradigm.

There are also tools in the market designed to recognize incorrect calculation logic after the model is built. Our application is designed to programmatically help in the initial model development, thereby "proactively" mitigating "human error" risks before they occur.

Finally, it is also important to note that empirical studies [Beckwith L., Cunha J., Fernandes J.P., Saraiva J., 2011] indicate model-based spreadsheets improve end user productivity. Model-Driven Engineering (MDE) techniques assist in end user effectiveness and efficiency. Therefore, benefits can be realized leveraging an MDE approach to spreadsheets.

Although the focus of the PivotModel is complex multi-dimensional modeling, a by-product of the application handles reporting and analysis. There is an extensive amount of Excel reporting tools in the marketplace which we don't consider ourselves as a replacement. As mentioned earlier, the intention of the PivotModel is to complement the Microsoft tool set. Therefore, our reporting and analysis is associated specifically with the PivotModel construct but the results of the PivotModel can be pushed to other reporting technologies if desired.

## 5. CONCLUSION

In this paper, we described our solution for mitigating risk in complex multi-dimensional models. This programmatic approach is unique and proactively addresses the issues of human error in spreadsheet model development.

There are tools in the marketplace designed to recognize incorrect calculation logic. These tools, however, work after a model is built. Our solution is designed to programmatically help in the initial model development. Consequently, the solution works to mitigate "human error" before it occurs.

Our solution aligns more with prior Model-Driven Engineering (MDE) concepts in that it approaches the problem by mitigating risk through design. However, these other alternatives begin with the premise of the Row, Column, Worksheet, Workbook paradigm as their baseline. Consequently, they approach the problem leveraging existing functionality like Range Names, Formula arrays, lookups, etc. to produce a resolution.

Although our solution may eventually leverage similar functions, it begins with the abstract concept of building a model outside the spreadsheet paradigm. Therefore, the framework of developing a solution entails three distinct components: Building a Model Structure, Creating Rule Logic and Loading Data. Once these components are defined, the solution dynamically generates the Excel-based data store called a PivotModel for users to slice and dice results.

Complex multi-dimensional solutions have been available for over 35 years. These types of applications require experienced developers with both a technical and business acumen. The challenge in bringing this technology to the desktop is figuring out ways in which to teach analysts how to utilize these tools. If we can create this new breed of analyst then we will see a transition from their current data cleansing and report creation role to one focused on data analysis and business performance.